\documentclass[review]{elsarticle}

\usepackage{hyperref}
\usepackage{graphicx}
\usepackage{amsmath}
\usepackage{amssymb}

\begin{document}

\begin{frontmatter}

\title{The aCORN Backscatter-Suppressed Beta Spectrometer}

\author[tul]{M.~T.~Hassan}
\author[nist]{F.~Bateman}
\author[ham]{B.~Collett}
\author[tul]{G.~Darius}
\author[tul]{C.~DeAngelis}
\author[nist]{M.~S.~Dewey}
\author[ham]{G.~L.~Jones}
\author[dep]{A.~Komives}
\author[tul]{A.~Laptev\fnref{LANL}}
\author[nist]{M.~P.~Mendenhall}
\author[nist]{J.~S.~Nico}
\author[ind]{G.~Noid}
\author[ind]{E.~J.~Stephenson}
\author[tul]{I.~Stern}
\author[tul]{C.~Trull}
\author[tul]{F.~E.~Wietfeldt}
\address[tul]{Department of Physics and Engineering Physics, Tulane University, New Orleans, LA 70118, USA}
\address[nist]{National Institute of Standards and Technology, Gaithersburg, MD 20899, USA}
\address[ham]{Physics Department, Hamilton College, Clinton, NY 13323, USA}
\address[dep]{Department of Physics and Astronomy, DePauw University, Greencastle, IN 46135, USA}
\address[ind]{CEEM, Indiana University, Bloomington, IN 47408, USA}
\fntext[LANL]{current address: Los Alamos National Laboratory, Los Alamos, NM 87545, USA }

\begin{abstract}
Backscatter of electrons from a beta spectrometer, with incomplete energy deposition, can lead to undesirable effects in many types of experiments. We present and discuss the design and operation of a backscatter-suppressed beta spectrometer that was developed as part of a program to measure the electron-antineutrino correlation coefficient in neutron beta decay (aCORN). An array of backscatter veto detectors surrounds a plastic scintillator beta energy detector. The spectrometer contains an axial magnetic field gradient, so electrons are efficiently admitted but have a low probability for escaping back through the entrance after backscattering. The design, construction, calibration, and performance of the spectrometer are discussed. 
\end{abstract}


\end{frontmatter}

\section{Introduction}
\label{S:Intro}
This beta spectrometer was developed as part of an experiment (aCORN) that measures the electron-antineutrino correlation $a$-coefficient in the beta decay of the free neutron. The most important features of neutron decay are described by the formula of Jackson, Treiman, and Wyld \cite{JTW57} which gives the neutron decay probability $N$ as a function of the neutron spin direction (\mbox{\boldmath $\sigma$}), and the momentum and total energy of the emitted electron (${\bf p_e}$, $E_e$) and antineutrino (\mbox{\boldmath $p_\nu$}, $E_{\nu}$):
\begin{equation}
\label{E:JTWeqn}
N \propto \frac{1}{\tau_n}E_e |{\bf p_e}| (Q-E_e)^2 \left[ 1 
+ a\frac{{\bf p_e}\cdot \mbox{\boldmath $p_\nu$}}{E_e E_\nu}
+ \mbox{\boldmath $\sigma$}\cdot \left( A\frac{\bf p_e}{E_e} + B\frac{\mbox{\boldmath $p_\nu$}}{E_\nu}
+ D\frac{({\bf p_e}\times \mbox{\boldmath $p_\nu$})}{E_e E_\nu} \right) \right].
\end{equation}
Here, $\tau_n$ is the neutron lifetime and $Q$ is 1293 keV, the neutron-proton rest energy difference. The parameters $a,A,B,D$ are correlation coefficients that are measured by experiments. In the Standard Model of the electroweak interaction these observables are closely related to fundamental parameters such as the axial vector and vector coupling constants $G_A$ and $G_V$ and the first element $V_{ud}$ of the CKM matrix. The $a$-coefficient determines the average angular correlation of ${\bf p_e}$ and \mbox{\boldmath $p_\nu$}. It is known to a relative precision of 4 \% from previous experiments \cite{Gri68,Str78,Byr02}: $a = -0.103\pm0.004$ \cite{PDG14}. 
An improved measurement of the $a$-coefficient may improve limits on scalar and tensor weak currents \cite{Dub91} and sharpen tests for possible conserved-vector-current (CVC) violation and second-class currents \cite{Gar01}.
\par
The aCORN experiment employs an asymmetry method first proposed by Yerozolimsky and Mostovoy \cite{Yer93, Bal94, Wie09}. It relies on a coincidence measurement of the beta electron and recoil proton. The electron energy and the time-of-flight (TOF) between electron and proton detection are measured. For each electron energy there are two groups of recoil protons, fast and slow, and the asymmetry in event rates of these is proportional to the $a$-coefficient. Unlike previous experiments, precise proton spectroscopy is not needed. The goal of aCORN is a measurement of the $a$-coefficient at the 1 \% relative uncertainty level, about a factor of five improvement over past experiments. Complete details of the experiment, its methods, goals, and analysis can be found in
\cite{Wie09,RSI16}.
\par
A key design feature of the aCORN beta spectrometer is suppression of backscattered electrons. Electrons that backscatter from the beta spectrometer without depositing their full energy can lead to a significant systematic effect because the electron energy is misidentified in such events. Backscattered electrons produce a low energy tail in the energy response function of the spectrometer. Based on Monte Carlo simulations of aCORN, we want this tail area to be $<$0.5 \% of the full-energy peak. A plastic scintillator detector will backscatter typically about 5 \% of incident electrons in the energy range 100 keV to 400 keV, so this implies a desired backscatter suppression efficiency of approximately \mbox{90 \%}.  Electron backscatter has long been recognized as a serious problem in beta spectroscopy using scintillation and solid-state detectors and various methods have been employed in the past to mitigate it \cite{Oke56,Gar58,Gri58,Chr72,Bop86,Mor93}. The particular needs of aCORN called for a novel solution.
\par
Additional design requirements for the aCORN beta spectrometer include:
\begin{enumerate}
\item{The beta electrons important to aCORN are in the energy range 100--400 keV and are selected by transverse momentum using a uniform axial magnetic field of about 36 mT and a series of collimating apertures with diameter 5.5 cm. These electrons must be accepted and counted by the beta spectrometer with high efficiency.}
\item{The electron energy response should be linear and measured with a calibration uncertainty of less than 2 \%.}
\item{The spectrometer must fit within the 64 cm space between the bottom of the electron collimator and the floor.}
\end{enumerate}
\section{Design and Construction}
\label{S:Design}
\begin{figure}[ht]
    \centering
    \includegraphics[width=4in]{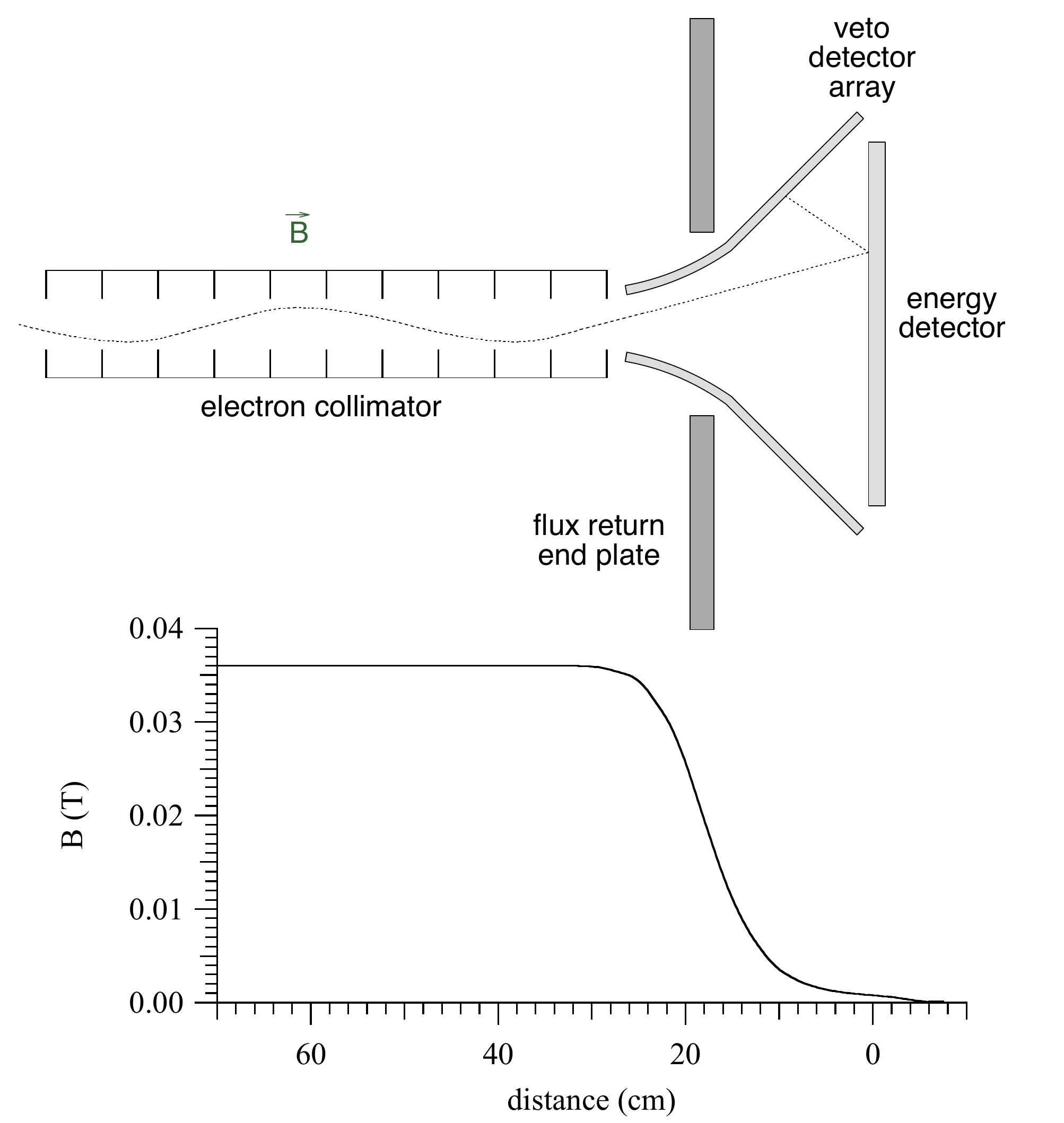}
    \caption{Top: conceptual design of the backscatter suppressed beta spectrometer. Electrons are transported (shown here from left to right as a dotted line) through the collimator in a uniform axial magnetic field, into the spectrometer and onto the energy detector. Of the electrons that backscatter, the majority will strike the veto detector array due to the weak magnetic field in that region.  Bottom: The calculated aCORN axial magnetic field, scaled and aligned to the top plot.}
    \label{F:bsscheme}
\end{figure}

Figure \ref{F:bsscheme} illustrates the design principles of the beta spectrometer. Beta electrons are transported from the decay region, with helical trajectories, through the electron collimator in a uniform axial 0.036 T magnetic field. Electrons with sufficiently small transverse momenta are accepted by the collimator and efficiently admitted into the beta spetrometer via the opening in the veto array. The axial magnetic field inside the spectrometer drops rapidly past the iron flux return plate with assistance from a set of additional trim coils. This drop in field causes electron trajectories to straighten and diverge from the axis. All accepted electrons with kinetic energy $>$100 keV strike the active area of the energy detector, a circular slab of plastic scintillator. Due to the shape of the magnetic field, when an electron backscatters from the energy detector the probability is low (about 3 \%) for it to be transported back through the entrance without striking the veto detector, an octagonal array of eight plastic scintillator paddles. A time coincidence of signals in the energy and veto detectors indicates a likely backscatter event which can then be removed from the data set. Prior to submission of the aCORN proposal, a prototype spectrometer based on this design concept was built and tested, and the results presented in a previous publication \cite{Wie05b}. The discussion here will focus on the second generation version that was used in the aCORN experiment.
\par
\begin{figure}[ht]
    \centering
    \includegraphics[width=6in]{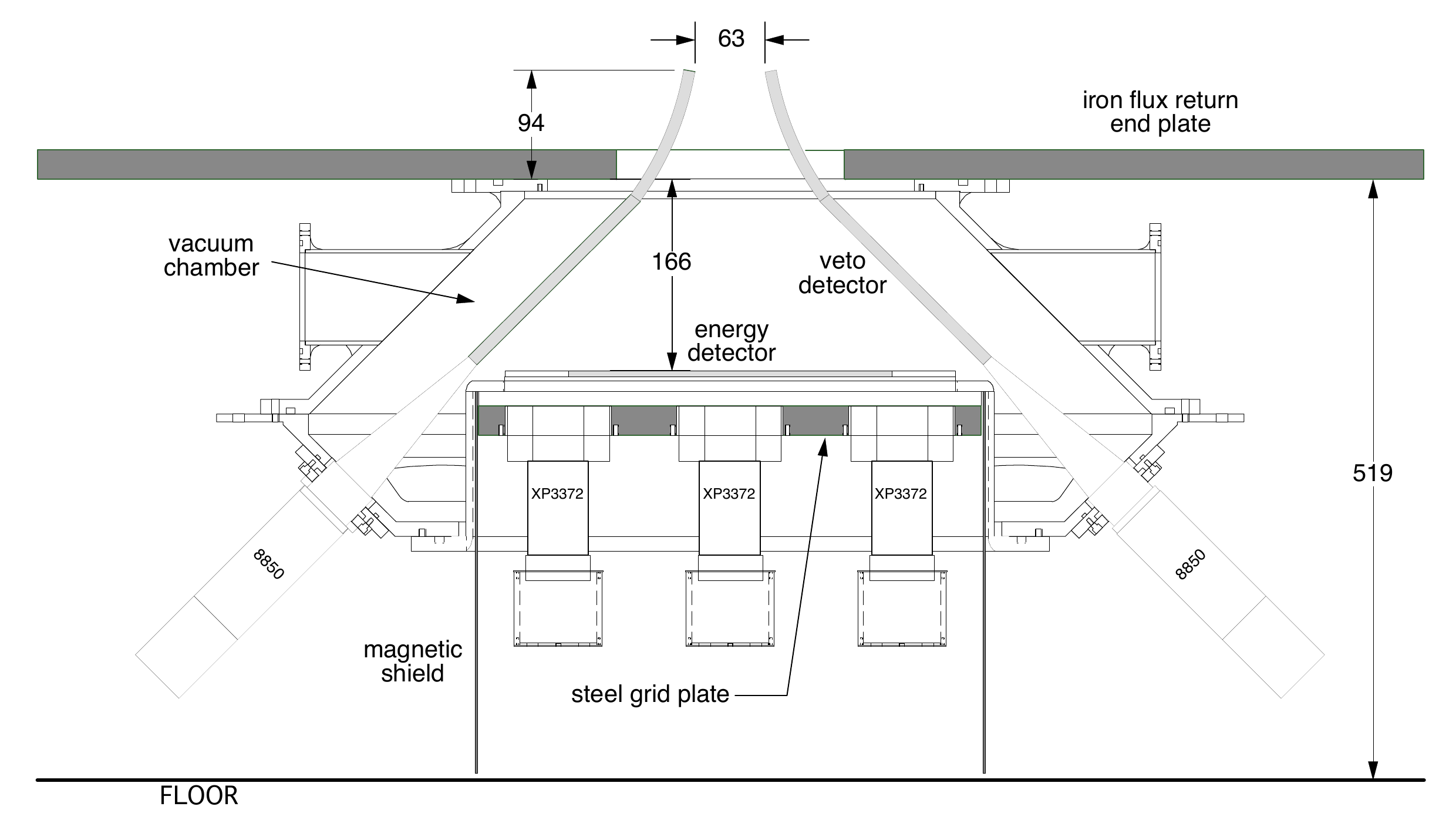}
    \caption{General arrangement of the spectrometer, shown mounted under the lower magnetic flux return plate of aCORN. Dimensions are in mm.}
    \label{F:bsCutaway}
\end{figure}
\begin{figure}[ht]
    \centering
    \includegraphics[width=3in]{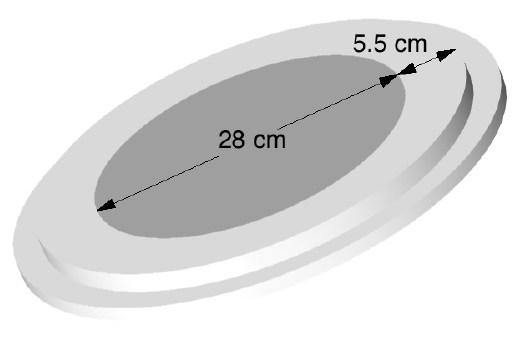}
    \caption{An illustration of the energy detector. Plastic scintillator (dark) is glued onto an acrylic light guide that also serves as the vacuum window. The 5.5 cm ring of acrylic around the scintillator reduces edge effects in the light collection efficiency.}
    \label{F:edet}
\end{figure}

The general arrangement of the spectrometer is shown in figure \ref{F:bsCutaway}. It is mounted to the underside of the aCORN apparatus, bolted to the bottom magnetic flux return end plate with a Viton o-ring vacuum seal. The energy detector is a phosphor-doped Bicron BC-408 polystyrene scintillator
\footnote{Certain trade names and company products are mentioned in the text or identified in an illustration in order to adequately specify
the experimental procedure and equipment used. In no case does such identification imply recommendation or endorsement by the National Institute of Standards and
Technology, nor does it imply that the products are necessarily the best available for the purpose.}. 
It is a 5 mm thick, 280 mm diameter circular slab. It is glued, using optical grade epoxy, to a 25 mm thick circular acrylic light guide that also serves as the vacuum window for the scintillation light. An additional acrylic ring, 5 mm thick and 55 mm wide, is glued surrounding the scintillator and flush with it. The ring serves to minimize edge effects in the light collection efficiency, ensuring that all points in the scintillator are viewed by a similar arrangement of photomultiplier tubes (PMTs), to improve the energy resolution. The configuration of the energy detector is illustrated in 
figure \ref{F:edet}. Directly beneath the acrylic window, outside the vacuum, is the grid plate, a 25 mm thick circular steel plate containing 19 hexagonal cutouts in a honeycomb pattern. The grid plate has three functions: 1) it provides a strong rigid support for the atmospheric pressure force on the acrylic window; 2) it positions the 19 energy PMTs; and 3) it comprises part of the PMT magnetic shield. The scintillation light is collected by 19 7.6-cm (3 inch) hexagonal, 8-stage, Photonis XP3372 PMTs. Each has a resistor voltage divider base with a nominal applied voltage of -1280 V. A thin cylindrical iron magnetic shield extends from the grid plate to the floor.
\par
\begin{figure}[ht]
    \centering
    \includegraphics[width=4in]{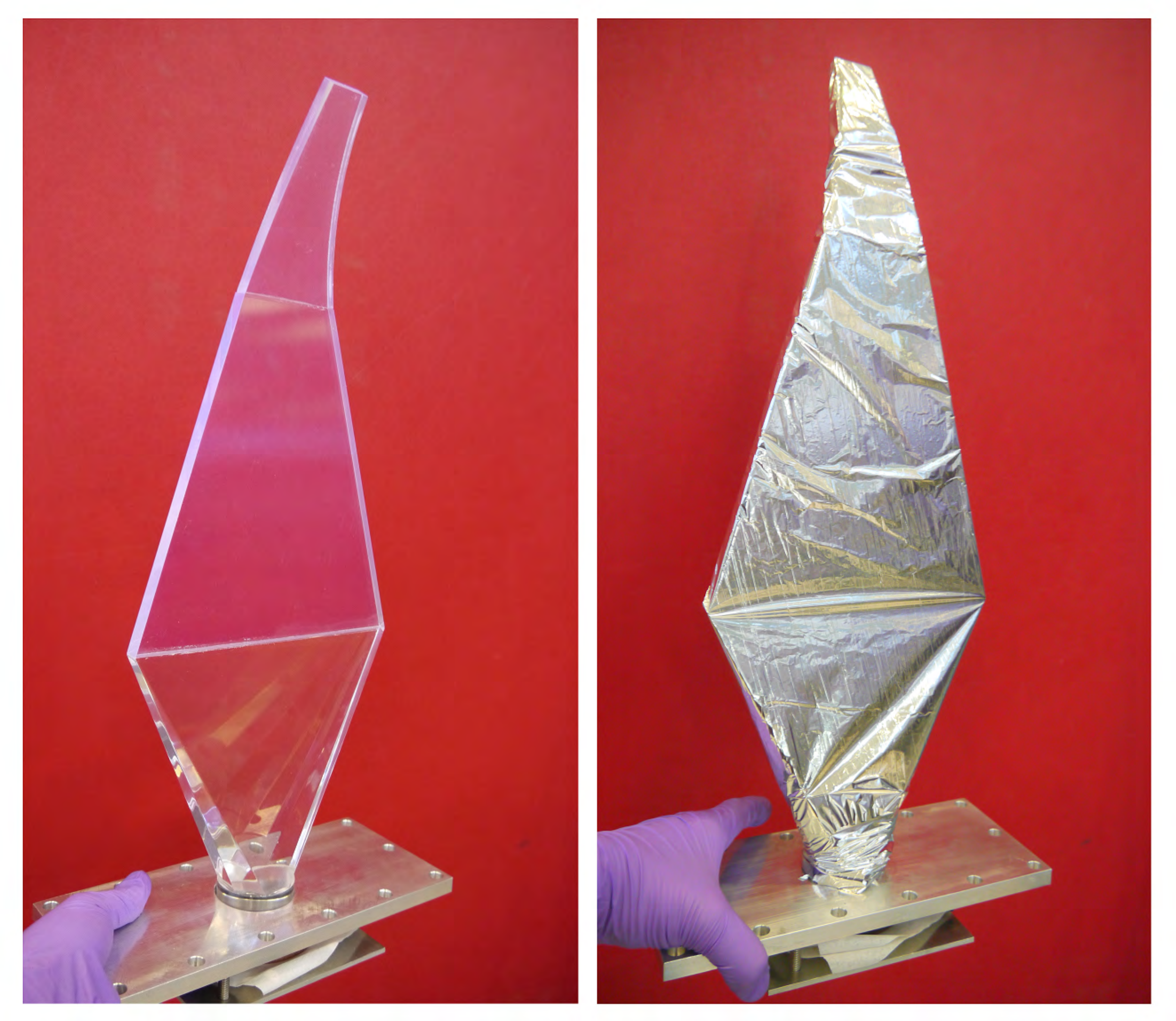}
    \caption{One of the eight veto detector paddles. Two trapezoidal sections of plastic scintillator, one curved and one flat, are glued to an adiabatic acrylic light guide. The photo on the right includes the mylar wrap for optical decoupling.}
    \label{F:vetopaddle}
\end{figure}

A closed octagonal array of eight veto detector paddles surrounds the energy detector. On each paddle the active detector region consists of two trapezoidal sections, one curved and the other flat, glued together. Both sections are 10 mm thick BC-408 plastic scintillator. The end of the lower section is glued to an ultraviolet-transmitting acrylic light guide that transforms the thin rectangular cross section of the scintillator to a circle adiabatically, {\em i.e.} the cross section varies gradually and continuously with increasing area for optimal light collection. At the bottom of the light guide is a vacuum penetration with an o-ring bayonet seal, below which it is coupled using optical grease to a 5.1-cm (2 inch) circular 12-stage Burle 8850 PMT. The Burle 8850 has an extremely high gain gallium phosphate first dynode so it is sensitive to weak scintillation light signals, providing excellent efficiency for detecting low energy backscattered electrons that strike any point of the active veto detector. Each veto PMT has a resistor voltage divider base with a nominal applied voltage of -2200 V. The scintillator and light guide of each veto paddle is wrapped in a single layer of thin aluminized mylar for optical decoupling from the energy detector. Figure \ref{F:vetopaddle} shows a complete veto paddle with and without the mylar wrapping. 
\par
\begin{figure}[ht]
    \centering
    \includegraphics[width=4in]{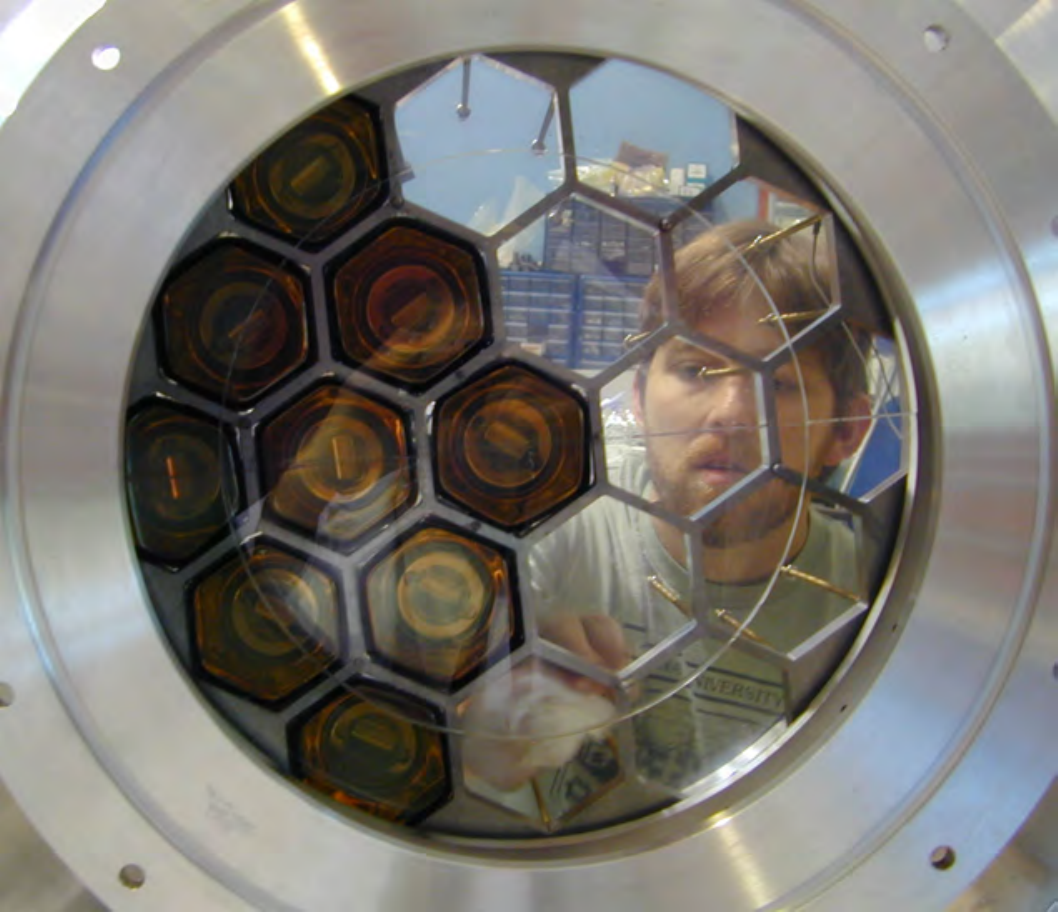}
    \caption{A view through the vacuum chamber of the energy detector, with the veto detector array removed, and 9 out of 19 energy PMTs installed.}
    \label{F:bsphoto1}
\end{figure}
\begin{figure}[ht]
    \centering
    \includegraphics[width=4in]{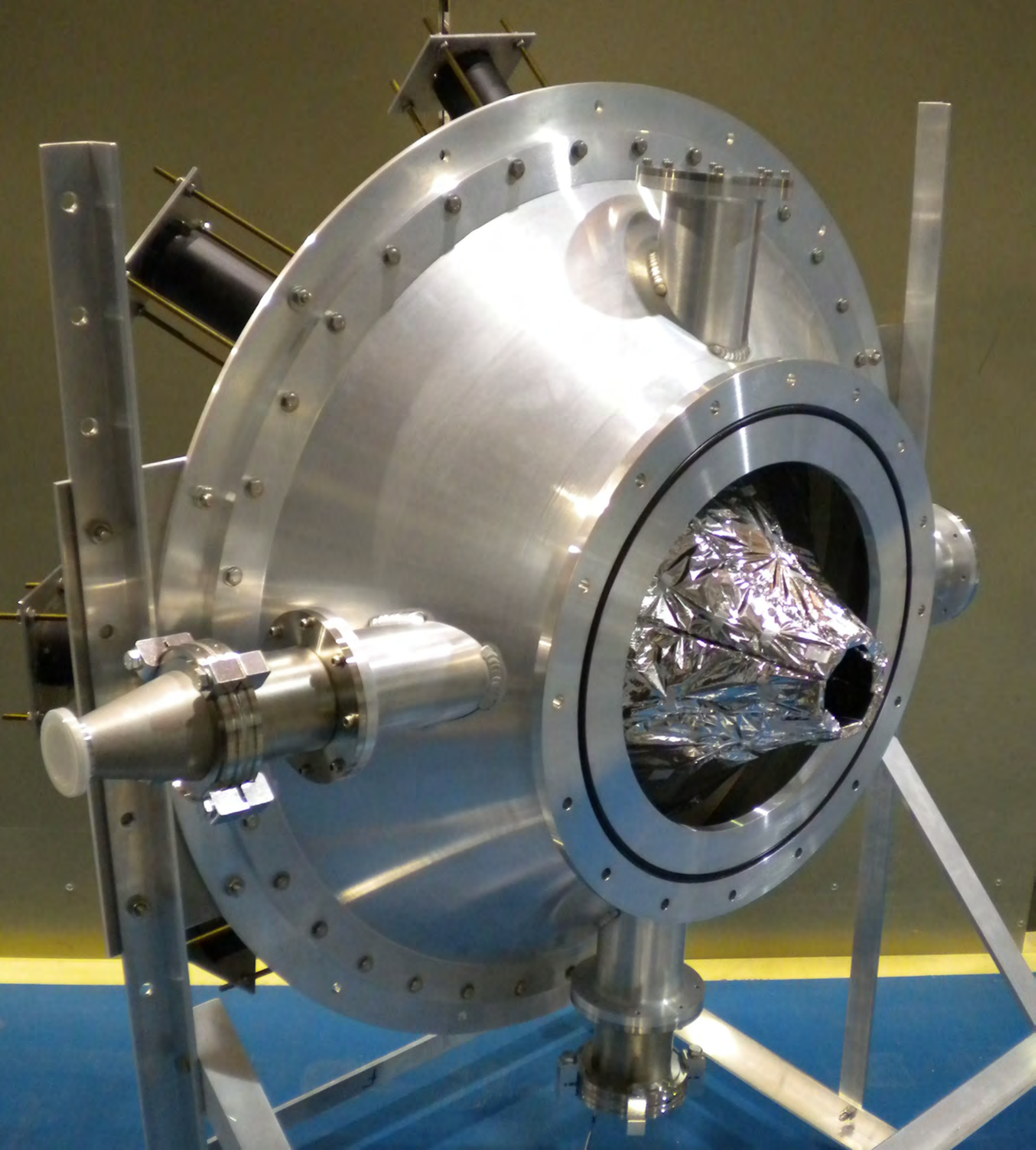}
    \caption{The complete beta spectrometer mounted horizontally on a test stand. The veto detector array is seen penetrating the entrance flange.}
    \label{F:bsphoto2}
\end{figure}
\begin{figure}[ht]
    \centering
    \includegraphics[width=4in]{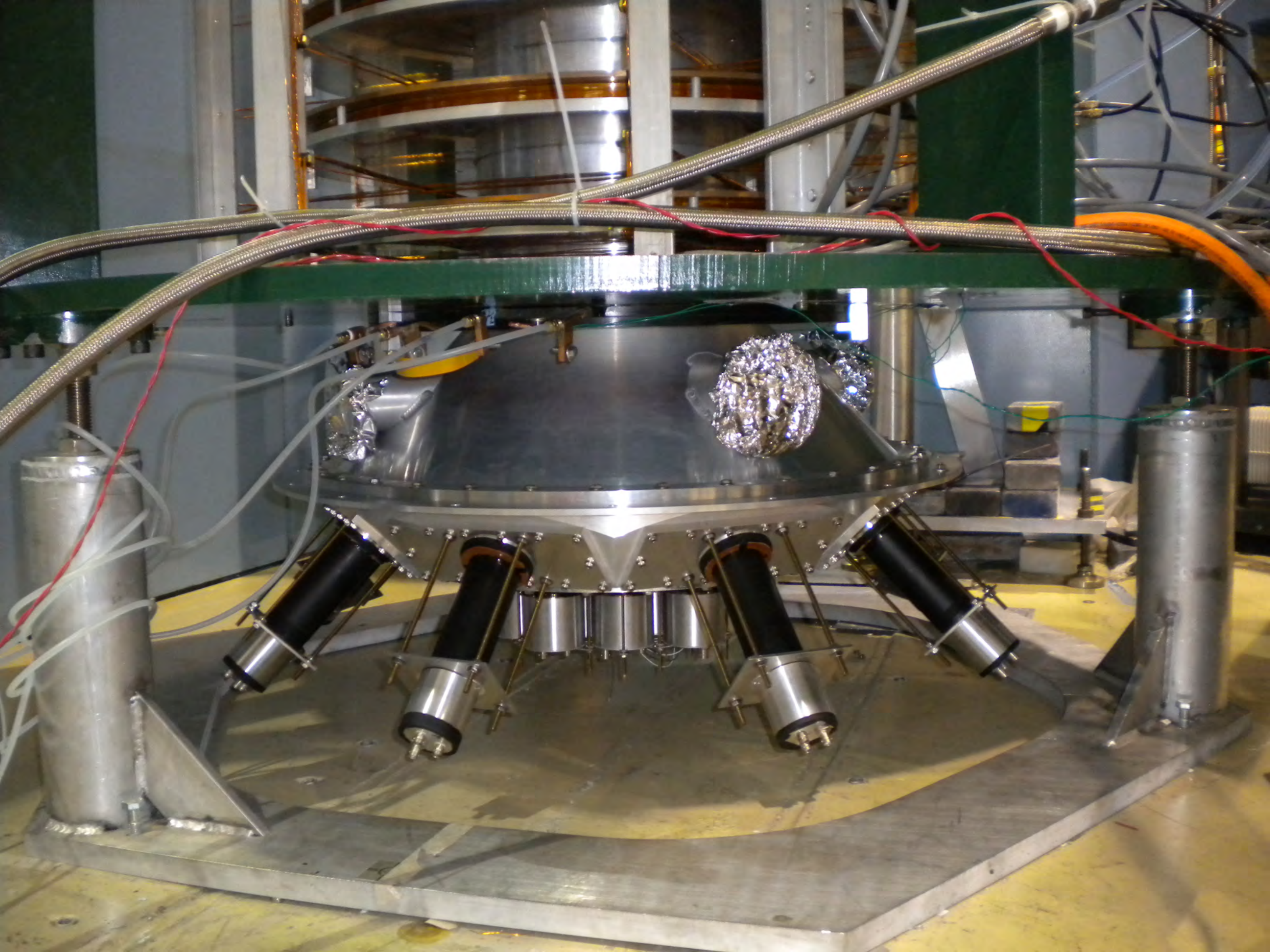}
    \caption{A photograph of the beta spectrometer installed in the aCORN experiment.}
    \label{F:bsphoto3}
    \end{figure}

The vacuum chamber is composed of three main sections, an upper shell, a lower shell, and the bottom flange that holds the energy detector assembly, all 6061 aluminum, bolted together with o-ring seals. Four large (10 cm) ports are provided for pumping and instrumentation. Figure \ref{F:bsphoto1} is a view of the energy detector in place. The scintillator/light guide assembly, grid plate, and PMT arrangement can be seen. Figure \ref{F:bsphoto2} shows the complete spectrometer mounted horizontally on a test stand, with a view of the veto paddle array. Figure \ref{F:bsphoto3} is a photograph of the spectrometer installed on the aCORN apparatus, beneath the magnetic flux return end plate.
\par
The energy and veto PMT high voltage is supplied by a computer-controlled 32 channel Wiener ISEG high voltage system. The data acquisition system (DAQ) is a 32 channel PIXIE-16 system (XIA LLC, Newark, CA), a 12 bit 100 MHz digitizer that accepts signals from the 19 energy and 8 veto PMTs and digitizes, integrates, and time-stamps each pulse. A minimum coincidence of any two PMT signals within 100 ns determines a raw event, for which energies and time stamps are written to a file for further processing.

\section{Calibration and performance}
\label{S:Cal}
\subsection{Conversion electron spectra}
\begin{figure}[ht]
    \centering
   \includegraphics[width=6in]{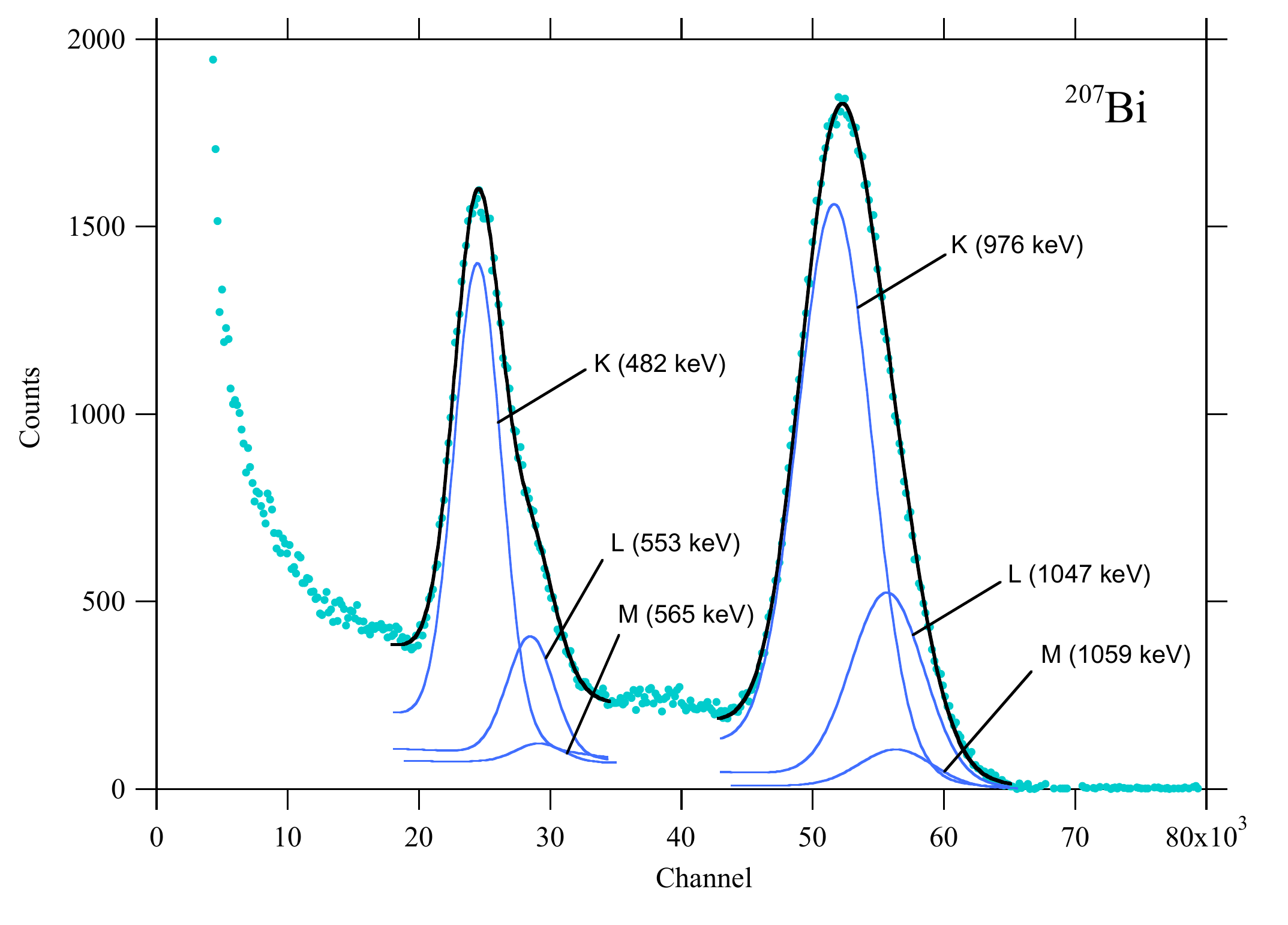}
    \caption{A $^{207}$Bi conversion electron calibration spectrum. 
    Each of the two main peaks were fit to a composite of the K, L, and M conversion line functions as described in the text. }
    \label{F:Bi207}
\end{figure}

\begin{figure}[ht]
    \centering
   \includegraphics[width=6in]{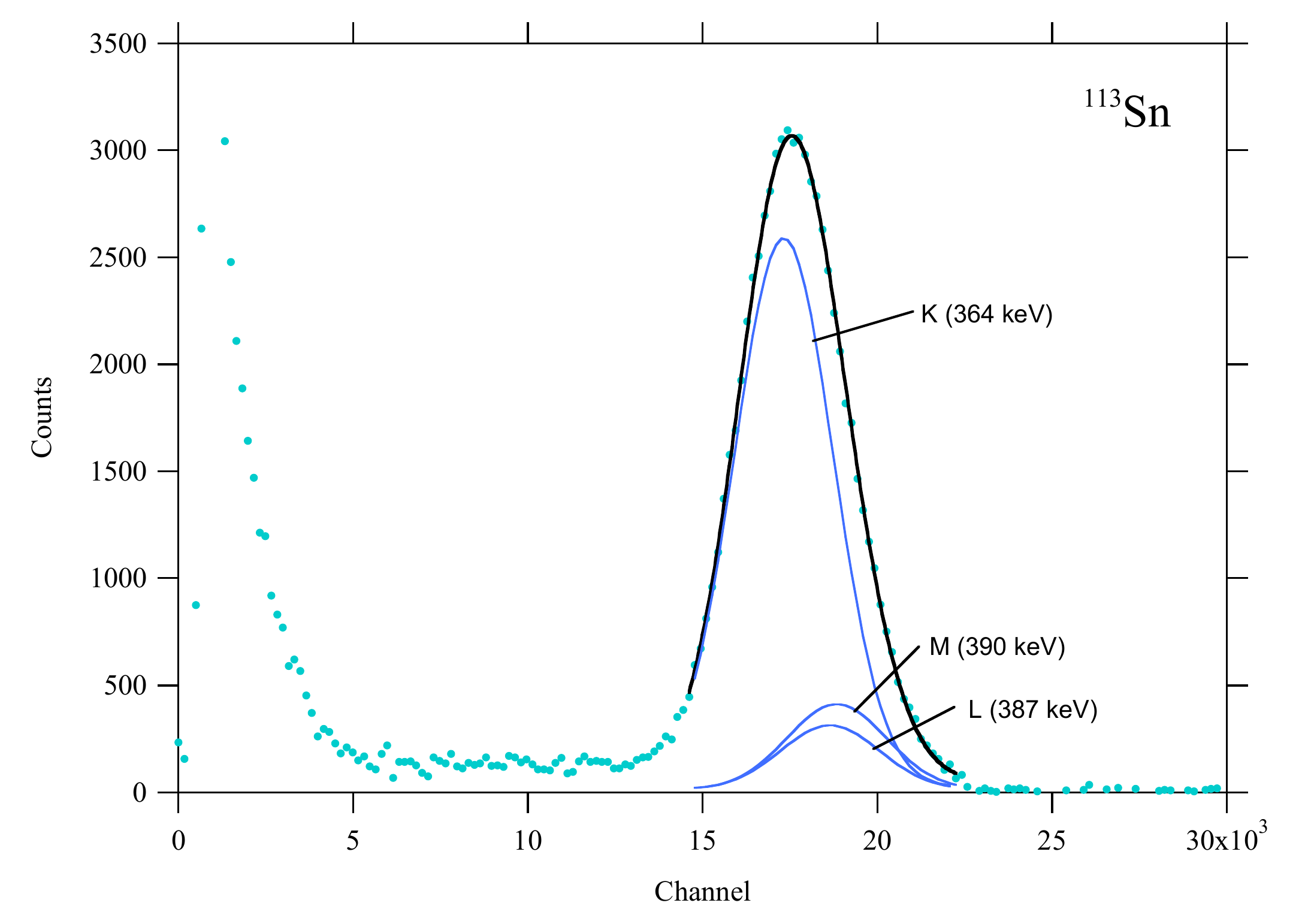}
    \caption{A $^{113}$Sn conversion electron calibration spectrum and composite fit. }
    \label{F:Sn113}
\end{figure}

Two test ports on the aCORN main vacuum chamber, located about 1 m above the beta spectrometer entrance, enabled {\em in situ} testing and calibration of the spectrometer. Ultrathin conversion electron sources, deposited on thin mylar substrates, were inserted periodically into position on the magnetic axis. Conversion electrons were transported efficiently by the magnetic field, while the distance provided effective suppression of the associated gamma and x rays. Sources used include $^{139}$Ce, $^{133}$Ba, $^{113}$Sn, and $^{207}$Bi with conversion electrons in the energy range 127 keV to 976 keV. Figure \ref{F:Bi207} shows a typical $^{207}$Bi spectrum. Each of the two main electron peaks was fit to a composite of three functions corresponding to the K, L, and M conversion lines for that transition. Each individual line function was represented by a Gaussian peak and a low energy tail approximated by an error function. The relative positions of the K, L, and M lines were fixed, but their individual intensities along with the overall Gaussian width and tail area were allowed to vary in the fit. A constant term was also included in each fit, for a total of seven fit parameters for each main peak. This procedure produced effective fits for each of the sources used. Figure \ref{F:Sn113} shows a similar fit to the $^{113}$Sn conversion electron peak.
\par
\begin{figure}[ht]
    \centering
   \includegraphics[width=5in]{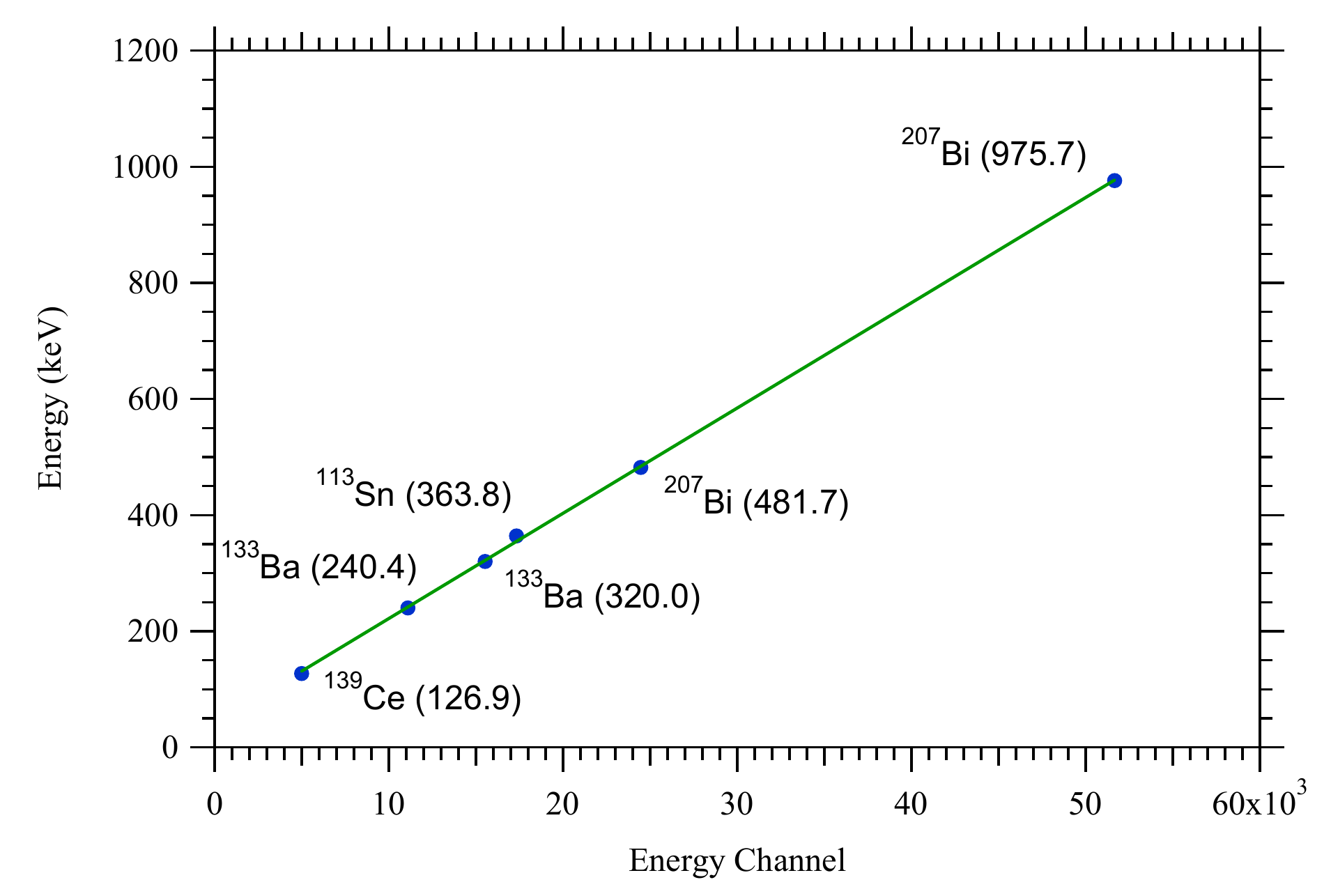}
   \caption{A linear fit (unweighted) of the K-conversion line energy (given in parentheses in keV) {\em vs.} the energy channel from each peak fit.}
    \label{F:energycal}
\end{figure}
\begin{figure}[ht]
    \centering
   \includegraphics[width=5in]{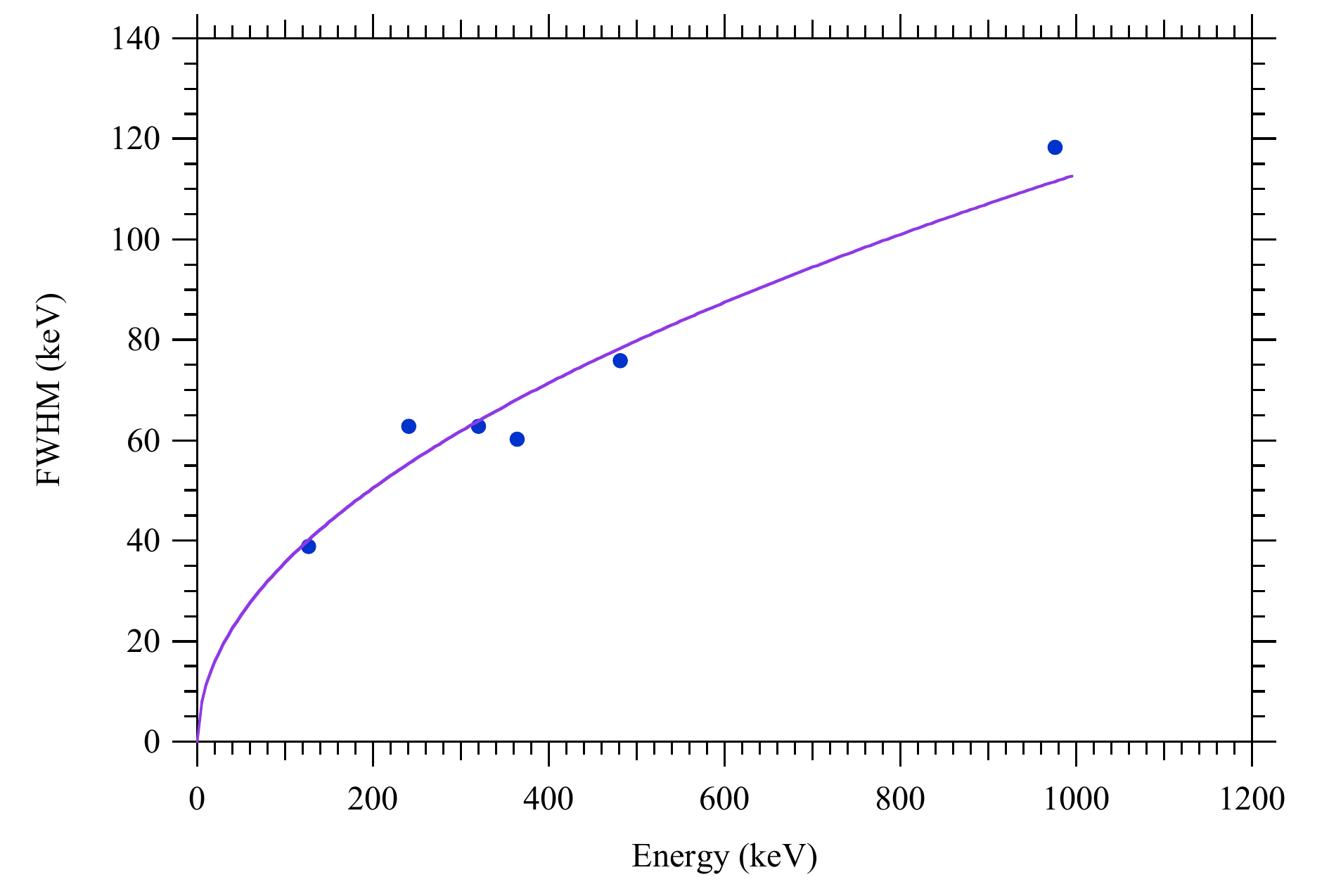}
    \caption{The full width at half maximum (FWHM) from the fit of each K-conversion line as a function of energy. The solid line is proportional to the square root of energy.}
    \label{F:energyres}
\end{figure}
\begin{figure}[ht]
    \centering
   \includegraphics[width=6in]{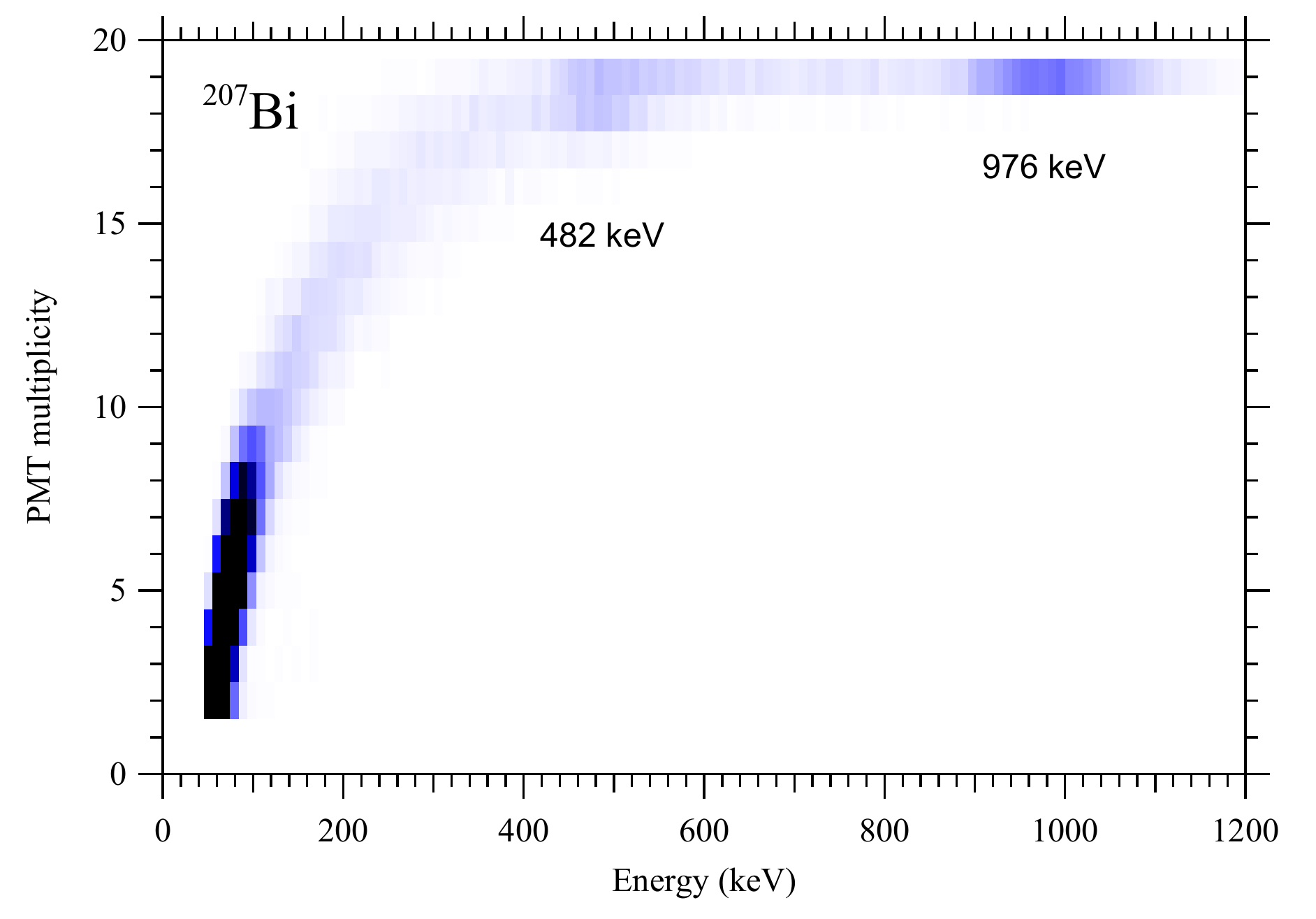}
    \caption{The energy-multiplicity histogram for a $^{207}$Bi conversion electron spectrum. The left axis is the number of energy PMTs (minimum 2 and maximum 19) that contributed to the event.}
    \label{F:EMplot}
\end{figure}

Figure \ref{F:energycal} is a linear fit of K-conversion peak energy {\em vs.} energy channel, which is the digital sum of all energy PMT signals determined by the DAQ to be associated with the event. The spectrometer exhibits excellent linearity in the energy range of interest. Figure \ref{F:energyres} shows full width at half maximum (FWHM) of the K-conversion line from each fit, as a function of energy. The FWHM is approximately proportional to the square root of energy, indicating that the energy resolution of the spectrometer is limited by photoelectron counting statistics, as desired. An interesting and useful diagnostic plot, the energy-multiplicity plot, is presented in figure \ref{F:EMplot}. It is a 2D histogram of energy PMT multiplicity as a function of event energy, in this case for a $^{207}$Bi spectrum. All 19 PMTs participated in events in the high energy peak, while the low energy peak contains multiplicity 16 to 19. When the spectrometer is functioning normally, all electron events regardless of their source will lie on this characteristic curve. Deviations can be used to identify and diagnose problems, such as a faulty PMT or DAQ malfunction.
\subsection{Neutron decay data}
\begin{figure}[ht]
    \centering
   \includegraphics[width=5in]{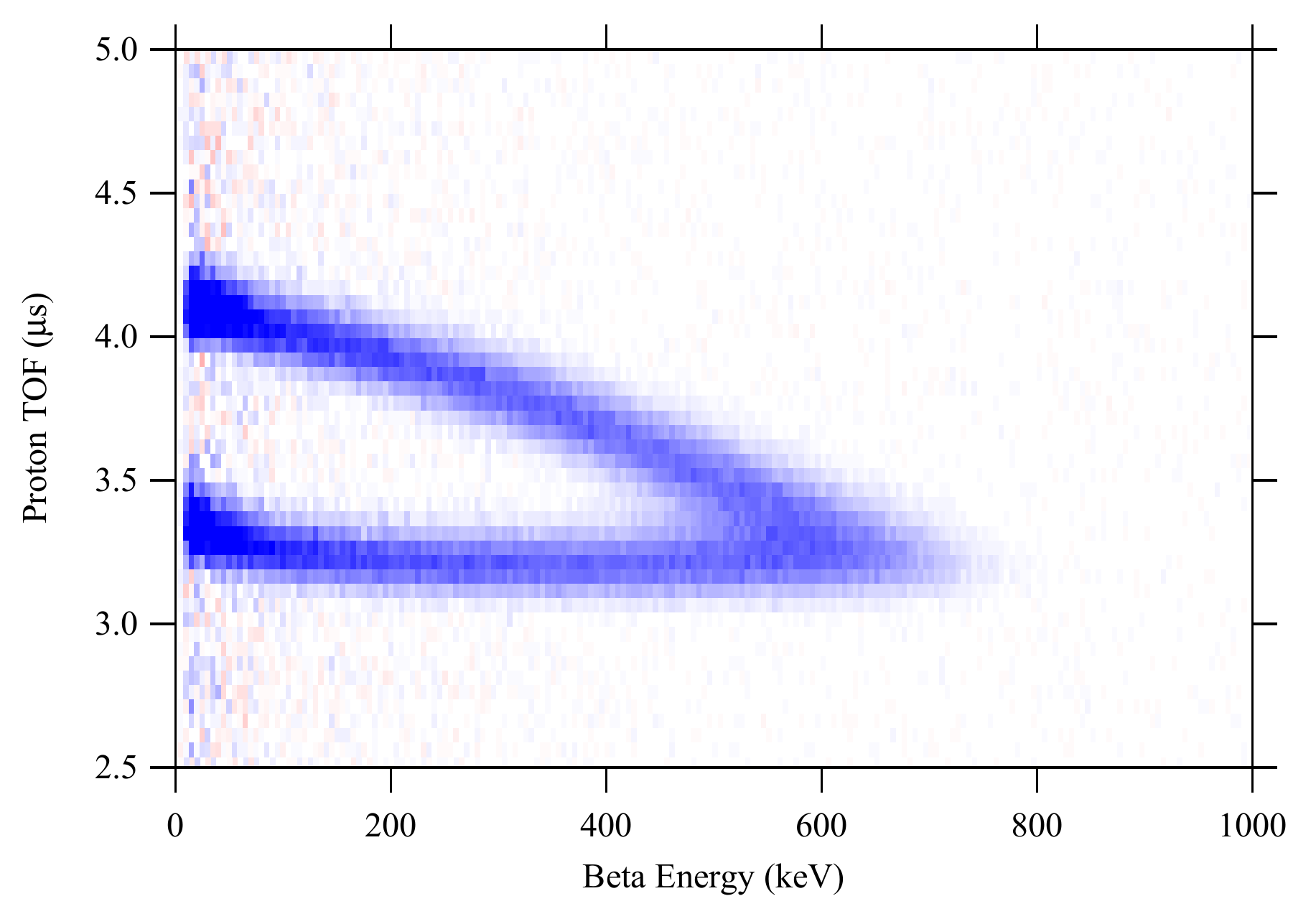}
    \caption{The background-subtracted aCORN ``wishbone'', a 2D histogram of proton time-of-flight (TOF) {\em vs.} beta energy for neutron decay coincidence events.}
    \label{F:wishbone}
\end{figure}

The aCORN experiment collected approximately two thousand hours of neutron decay data on the NG-6 beamline at the National Institute of Standards and Technology Center for Neutron Research in Gaithersburg, Maryland, USA. Figure \ref{F:wishbone} shows a typical 2D histogram, background subtracted,  of proton TOF {\em vs.} beta energy for events where a beta electron and recoil proton were detected in coincidence. It forms a characteristic ``wishbone'' shape. For beta energies below about 400 keV, there are two distinct proton TOF groups, fast and slow. The asymmetry in counts of these groups is proportional to the $a$-coefficient. See \cite{Wie09,RSI16} for further discussion of the aCORN wishbone and its interpretation. Electron backscatter from the beta spectrometer, if not sufficiently suppressed, will cause events to appear at the wrong (lower) energy in the wishbone and tend to fill in the gap between the fast and slow proton groups, confounding the asymmetry determination.
\par
\begin{figure}[ht]
    \centering
   \includegraphics[width=5in]{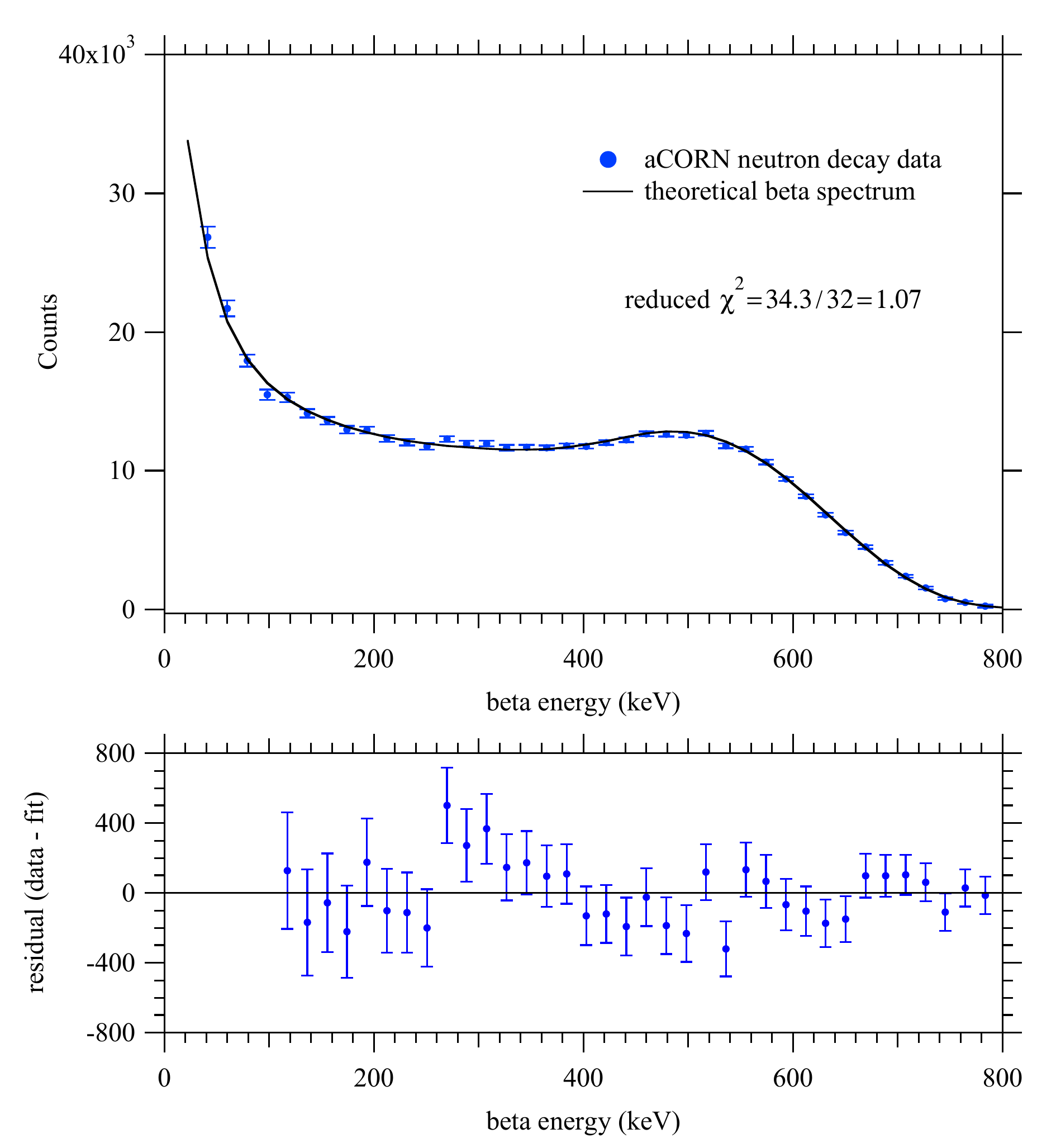}
    \caption{The wishbone energy spectrum fit to the calculated theoretical shape, which includes the aCORN momentum acceptance and a simple Gaussian resolution function.  Error bars are statistical uncertainties.}
    \label{F:wbspectrum}
\end{figure}

If we take the wishbone histogram and sum over TOF for each energy, we obtain the wishbone energy spectrum. This is essentially the Fermi beta decay energy spectrum, with its endpoint at 782 keV for neutron decay, modified by the aCORN transverse momentum acceptance of electrons and protons in coincidence. Figure \ref{F:wbspectrum} shows a wishbone energy spectrum fit to the calculated theoretical shape for aCORN. Here the theoretical spectrum has been convoluted with a Gaussian of width $\sigma = c \sqrt{E}$, where $E$ is beta kinetic energy and $c$ is a constant. This $\sqrt{E}$ dependence of the energy resolution is expected for a scintillator detector whose resolution is limited by photoelectron counting statistics. A good fit is obtained from 100--800 keV with four free parameters in the fit: the linear energy calibration slope and offset, an overall scale factor, and the energy resolution parameter $c$. In practice we found that this fit of the neutron decay wishbone energy spectrum provided the best energy calibration of the spectrometer for the experiment because: 1) there was no energy-dependent structure in the background that can cause systematic errors, unlike the conversion source spectra; 2) statistical uncertainties in the calibration slope and offset were smaller;  and 3) it avoided problems due to a rate-dependent calibration offset that was observed.
\subsection{Backscatter suppression efficiency}
\begin{figure}[ht]
    \centering
   \includegraphics[width=5in]{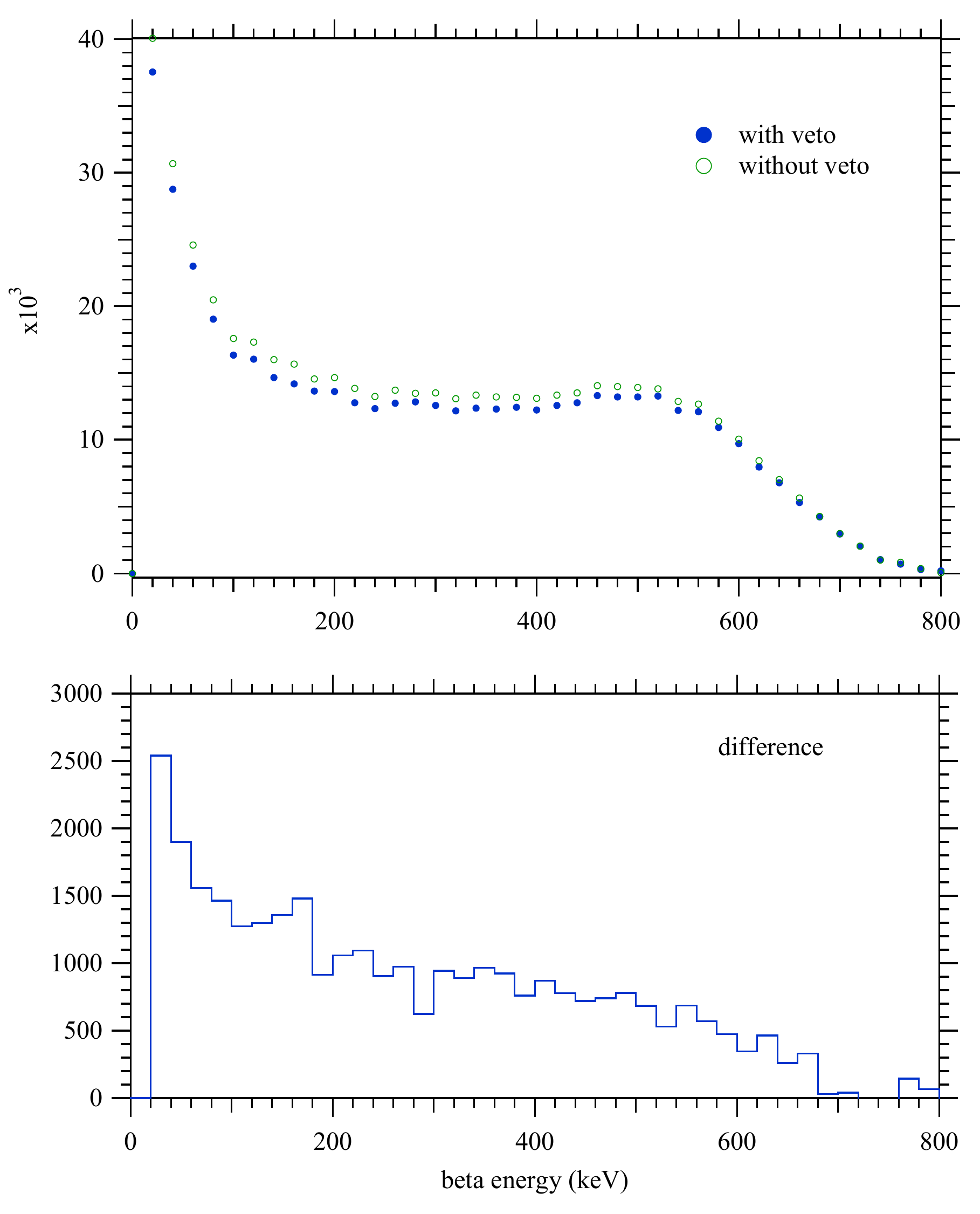}
    \caption{Top: Solid circles are the wishbone energy spectrum with the backscatter veto in effect. Open circles are the
    equivalent spectrum but without the backscatter veto. Bottom: The difference spectrum (without veto minus vetoed) containing neutron decay events where the electron  backscattered.}
    \label{F:vetocompare}
\end{figure}

Based on its design we estimate the backscatter suppression efficiency of the spectrometer at approximately 90 \%. Directly measuring it is a challenge. Our best determination is from the neutron decay data which, unlike the conversion electron data, is free of source scattering that contributes significantly to a tail in the response function. Figure \ref{F:vetocompare} (top) shows a comparison of the wishbone energy spectrum with and without events that produced a signal in the veto array. The bottom figure is the difference spectrum, the electrons that backscattered from the energy detector and then struck one of the veto paddles. In the energy range 100 keV to 780 keV it contains \mbox{6.1 \%} of the unvetoed spectrum. To calculate the backscatter suppression efficiency, we need to compare this to the number of electrons that backscattered but were not vetoed. Our best measure of that is found in the kinematically forbidden gap between the fast and slow proton branches of the wishbone histogram (figure \ref{F:wishbone}). A careful comparison of the Monte Carlo generated wishbone with aCORN data gives a result that is statistically consistent with zero.  There is no evidence of neutron decay events in the gap, which would indicate the presence of an unsuppressed backscatter tail, but the statistical fluctuation of counts in that region due to the background subtraction is significant and results in a one sigma upper bound of 1.2 \%.  There is another potential source of gap events: electrons that scattered and lost energy in the electron collimator and were subsequently detected. The collimator was carefully designed to minimize such events, but a Monte Carlo analysis predicts a tail contribution of about 0.3 \% from them. Finally, a Monte Carlo simulation of the beta spectrometer showed that 3 \% of backscattered electrons will escape through the opening in the veto array. We obtain from these considerations a probable backscatter suppression efficiency in the range 87--97 \%, in good agreement with expectation.

\section{Summary}
We designed, constructed, and tested a plastic scintillator beta spectrometer. An array of plastic scintillator veto detectors, in concert with a specifically tailored magnetic field profile, enables suppression of electron backscatter from the primary energy detector with an efficiency of approximately 90 \%. The measured energy response in the range 100 keV to 1 MeV is linear and calibrated with relative uncertainty $<$1 \%. The energy resolution (FWHM) at 500 keV is 16 \%, with an energy dependence proportional to square root of energy. This spectrometer was used successfully for several years as part of the aCORN neutron decay correlation experiment at the NIST Center for Neutron Research. 

\section{Acknowledgments}
\label{S:Ack}
This work was supported by the National Science Foundation, U.S. Department of Energy Office of Science, and NIST (US Department of Commerce). We thank the NIST Center for Neutron Research for technical support and for providing the neutron facilities used in this work. 

\bibliography{aCORNbsv4X}

\end{document}